\documentclass[letterpaper,10pt]{article}
\usepackage[T1]{fontenc}
\usepackage[utf8]{inputenc}
\usepackage{parskip}
\usepackage[font={small}]{caption}
\usepackage[bookmarks=false,hidelinks]{hyperref}
\usepackage[margin=2.3cm,top=1.4cm,bottom=2cm]{geometry}
\usepackage{times}
\usepackage{amsmath,amssymb,amsthm}
\usepackage[affil-it]{authblk}
\usepackage{listings}
\usepackage[round]{natbib}
\usepackage{graphicx}

\date{}

\DeclareMathOperator*\argmax{arg\,max}

\usepackage{amsfonts}
\usepackage[ruled,vlined,linesnumbered]{algorithm2e}

\SetCommentSty{mycommentfont}
\DontPrintSemicolon
\SetKwComment{mycomment}{}{}

\usepackage{enumerate}

\usepackage{xcolor}

\usepackage[font={small}]{caption}
\usepackage[font={small}]{subcaption}

\usepackage{pgfplots}
\pgfplotsset{compat=newest}
\usetikzlibrary{plotmarks}
\usetikzlibrary{arrows.meta}
\usepgfplotslibrary{patchplots}

\newlength\figureheight
\newlength\figurewidth

\usepackage{graphicx}      

\renewcommand\mid{\,\vert\,}

\newcommand{\nii}{{n}}
\newcommand{\nij}{{j}}

\newcommand{\Prb}[1]{\mathbb{P}\left({#1}\right)}
\newcommand{\Exp}[1]{\mathbb{E}\left[{#1}\right]}

\newcommand{\param}{\theta}

\newcommand{\Reals}{\mathbb{R}}
\newcommand{\N}{\mathcal{N}}
\newcommand{\pss}[1]{\{#1\}_{^{n=1}}^{_{N}}}

\newcommand{\MLparam}{\smash{\widehat{\param}}}
\newcommand{\lik}{p_\param(y_{1:T})}
\newcommand{\likest}{\smash{\widehat{z}_\param}}
\newcommand{\likestn}[1]{\smash{\widehat{z}_{#1}(\param)}}

\title{Learning Nonlinear State-Space Models Using Smooth Particle-Filter-Based Likelihood Approximations}
\author{Andreas Svensson\thanks{\url{andreas.svensson@it.uu.se}}}
\author{Fredrik Lindsten}
\author{Thomas B. Schön}
\affil{Department of Information Technology, Uppsala University}

\begin{document}

\maketitle

%
%
%

\begin{abstract}
When classical particle filtering algorithms are used for maximum likelihood parameter estimation in nonlinear state-space models, a key challenge is that estimates of the likelihood function and its derivatives are inherently noisy. The key idea in this paper is to run a particle filter based on a current parameter estimate, but then use the output from this particle filter to re-evaluate the likelihood function approximation also for other parameter values. This results in a (local) deterministic approximation of the likelihood and any standard optimization routine can be applied to find the maximum of this local approximation. By iterating this procedure we eventually arrive at a final parameter estimate.
\end{abstract}



\section{Introduction}

Consider the nonlinear state-space model
\begin{subequations}\label{eq:ssm}
	\begin{align}
	x_{t}\mid x_{t-1} &\sim f_\param(x_{t}\mid x_{t-1}),\\
	y_{t}\mid x_t &\sim g_\param(y_{t}\mid x_t),
	\end{align}
\end{subequations}
where the hidden states $x_t \in \mathsf{X} \subset \Reals^{N}$ and observations $y_t \in \mathsf{Y} \subset \Reals^{n_y}$ evolves over time $t = 0, 1, \dots$. We assume that the probability densities $f_\param$ and $g_\param$ are parameterized by unknown parameters $\param$, which we are interested to learn from data. More specifically we are seeking the maximum likelihood estimate of $\param$ from given observations $\{y_1, \dots, y_T\} \triangleq y_{1:T}$, i.e., to solve
\begin{equation}
\MLparam = \argmax_\param \lik.\label{eq:ml}
\end{equation}
We shall refer to $\lik$ as the \emph{likelihood function} when considered a function of $\param$. It follows from~\eqref{eq:ssm} and the initial state density $p(x_0)$ that
\begin{equation}
\lik = 
\int p(x_0)\prod_{t=1}^{T}f_\param(x_{t}\mid x_{t-1}) g_\param(y_t\mid x_t)dx_{0:T}.\label{eq:lik}
\end{equation}
This can be evaluated analytically only for a few special cases. Approximations, such as the particle filter, are needed for the nonlinear case. For any given $\param$, the particle filter can be run to give a Monte Carlo based estimate $\likest$ of~\eqref{eq:lik}. It is well-known \citep{DelMoral:2004} that $\likest$ is unbiased, i.e., $\Exp{\likest} = \lik$, where the expectation is over the stochasticity in the particle filter algorithm itself. Albeit unbiased, $\likest$ is also stochastic (i.e., a different value is obtained every time the particle filter algorithm is run, since it is a Monte Carlo solution), often with a considerable variance and non-Gaussian characteristics, and standard optimization routines can therefore not be used to solve~\eqref{eq:ml}. To this end, schemes using, 
optimization for noisy function evaluations have been applied (e.g., \citealt{WS:2017,DL:2014}).

We propose in this paper to iteratively run a particle filter with some current parameter estimate $\param_{k-1}$ and scrutinize the particle filter algorithm to construct a \emph{deterministic}\\local approximation of the likelihood function around $\param_{k-1}$. In a sense we hereby manage to circumvent the stochasticity from the Monte Carlo procedure in a consistent way, and can allow for standard optimization routines to search for a new parameter value $\param_{k}$, for which we can run a new particle filter, etc. The idea is outlined as Algorithm~\ref{alg:idea}. Related ideas have been proposed earlier \citep{LeGland:2007},
but they have to the best of the authors knowledge not yet been fully explored in the system identification context.

\begin{algorithm}[h!]
	\For{$k = 1, \dots$}{
		Run a particle filter with $\param_{k-1}$ and save all the generated random numbers $\pss{x^\nii_{0:T},a^\nii_{1:T}}$.\;
		Re-write the likelihood estimator $\likest$ as a deterministic function of the particle system $\pss{x^\nii_{0:T},a^\nii_{1:T}}$, and use conventional optimization to find its maximizing argument $\param_k$.\;
	}
	\caption{Identification idea}
	\label{alg:idea}
\end{algorithm}

\section{Background on particle filtering}

The particle filter was originally proposed as a Monte Carlo solution to the state filtering problem \citep{GSS:1993}, i.e., to compute $p_\param(x_t\mid y_{1:t})$. It was soon \citep{Kitagawa:1996}  realized that the particle filter could also be used to \emph{estimate the likelihood function} for given parameter values, essentially by inserting the particles (Monte Carlo samples) into the integral in \eqref{eq:lik} and thereby obtain an estimate $\likest$ of $\lik$. Thanks to this likelihood estimate, the particle filter can be used for system identification purposes.\footnote{There are several alternative ways in which the particle filter can be used for system identification, for example approaches based on the EM algorithm or Gibbs sampling.} As mentioned, $\likest$ is unbiased, but it often has a heavy-tailed and asymmetric distribution with a non-negligible variance. Its exact properties depends on the particle filter settings and the model.

A textbook introduction to particle filters is given in, e.g., \citet{SL:2016, DFG:2001}. We summarize a rather general formulation of the particle filter in Algorithm~\ref{alg:pf}, a version of the auxiliary article filter \citep{PS:1999}. We use $q(x_t\mid x_{t-1},y_t)$ to denote an almost arbitrary\footnote{The support of $q$ has to cover the support of $f_\param$.} proposal distribution. Furthermore, $\nu_t^{_\nii}$ are the resampling weights, $w_t^{_\nii}$ the importance weights, and $\mathcal{C}(\pss{\nu_t^{_\nii}})$ denotes the categorical distribution on the set $\{1, \dots, N\}$ with (possibly unnormalized) weights $\pss{\nu_t^{_\nii}}$, and $N$ is the number of particles.

Let us be explicit about the stochastic elements of Algorithm~\ref{alg:pf}: Particles are initially drawn from the initial distribution $p(x_0)$ on line \ref{alg:pf:i} (which we assume to be independent of $\param$). Furthermore, the ancestor indices $a_t^{_\nii}$ on line~\ref{alg:pf:res} are drawn with respect to the resampling weights $\nu_t^{_\nii}$, and finally the propagation of particles $x_t^{_\nii}$ on line~\ref{alg:pf:prop}, where the new particles are drawn from the proposal $q(x_t\mid x_{t-1},y_t)$.

Whereas $f_\param(x_t\mid x_{t-1})$, $g_\param(y_t \mid x_t)$ and $p(x_0)$ in Algorithm~\ref{alg:pf} are given by the model specification and $w_t^{_\nii}$ follows from the algorithm, the choices for $q(x_t\mid x_{t-1}, y_t)$, $\nu_t^{_\nii}$ and $N$ are left to the user. The number of particles $N$ is usually taken as large as the computational budget permits, and two common choices for weights and proposals are
\begin{itemize}
	\item the \emph{bootstrap particle filter} \citep{GSS:1993} with \[q(x_t\mid x_{t-1}, y_t) = f_\param(x_t\mid y_t)\] and \[\nu_t^\nii = w_t^\nii,\] and consequently $w_t^{_\nii} = g_\param(y_t\mid x_t^{_\nii})$. This choice is generic and requires very little from the user, but has inferior performance compared to
	\item the \emph{fully adapted particle filter} \citep{PS:1999}, with \[q(x_t\mid x_{t-1}, y_t) = p_\param(x_t\mid x_{t-1}, y_t)\] and \[\nu_t^\nii = p_\param(y_t\mid x_{t-1}).\] This choice is superior to the bootstrap choice in terms of variance of the obtained approximation, but it is only available for a quite limited set of models. The literature on approximations to this choice is therefore rich (e.g., \citealt{NLS:2015,DGA:2000}).
\end{itemize}

We will in this paper exploit the relatively large freedom that is available when it comes to choosing the proposal density and the resampling weights. 
By making a choice that only depends on the \emph{current} parameter value~$\param_{k-1}$ it is possible to evaluate the likelihood estimator (which we will denote by $\likestn{\param_{k-1}}$) for different values of $\param$, while at the same time making use of the \emph{same} realization of  $\pss{x^{_\nii}_{0:T},a^{_\nii}_{1:T}}$.

\begin{algorithm}[t]
	Draw ${x}_0^\nii \sim p({x}_0)$ and set $w_0^\nii = 1, \nu_0^\nii = 1$.\label{alg:pf:i}\;
	\For{$t = 1$ \KwTo $T$}{
		Draw ${a_t^\nii} \sim \mathcal{C}(\{\nu_{t-1}^\nij\}_{j=1}^N).$\label{alg:pf:res}\;
		Propagate ${x^{\nii}_{t}} \sim q({x_{t}} \mid {x_{t-1}^{a_t^\nii}}, y_t)$.\label{alg:pf:prop}\;
		Set	$w_t^\nii \leftarrow \frac{w_{t-1}^{a_t^\nii}/\sum_{j=1}^{N}w^{j}_{t-1}}{\nu_{t-1}^{a_t^\nii}/\sum_{j=1}^N \nu_{t-1}^{a_t^\nij}}~\frac{f_{\theta}(x_t^\nii \mid x_{t-1}^{a_t^\nii})}{ q(x_t^\nii \mid x_{t-1}^{a_t^\nii}, y_t)}g_{\theta}(y_t \mid x_t^\nii)$.\;
		Set $z_t \leftarrow \frac{1}{N} \sum_{n=1}^{N} w_t^\nii$.\;
	}
	Compute $\likest = \prod_{t=1}^{T} z_t$.\;
	\BlankLine
	\mycomment{All statements with $n$ are for $n = 1, \dots, N$.}
	\caption{The auxiliary particle filter}
	\label{alg:pf}
\end{algorithm}

\section{Related work}

The use of the likelihood estimator $\likest$ as an objective function in optimization, in the search for a maximum likelihood estimate $\MLparam$, has been subject to several studies: \citet{DT:2003} differentiate $\likest$ and use it in a stochastic gradient descent scheme, whereas \citet{DL:2014,WS:2017} use an optimization scheme based on Gaussian processes. \citet{MP:2011} use a fixed random seed and run the particle filter for different $\param$. For a fixed random seed, $\likest$ is indeed deterministic, however with discontinuous `jumps' due to different resampling decisions being made for different $\param$. To this end, \citeauthor{MP:2011} proposes an approximative smoothing to obtain a continuous function.

The idea used in this paper, i.e., to make the ancestor indices $a^{_\nii}_{1:T}$ and particles $x^{_\nii}_{0:T}$ depending only on the current, or reference, parameter value $\param_{k-1}$ (instead of $\param$) has been analyzed by \citet{LeGland:2007}, but has (to the best of the authors knowledge) not yet been applied to the system identification problem. The work by \citeauthor{LeGland:2007} concerns central limit theorems for the likelihood estimator $\likestn{\param_{k-1}}$. The application of the results from \citet{LeGland:2007} to this work is subject for further work. The work presented in this paper shares similarities with the authors recent work \citet{SSL:2017}, which however is concerned with the different topic of Bayesian identification for state-space models with highly informative observations.

Other approaches to maximum likelihood estimation in nonlinear state-space models include the combination of the popular Expectation Maximization (EM) algorithm and particle filters \citep{Lindsten:2013,SWN:2011,ODC+:2008}.

\section{The solution}

The key idea in this paper is to choose $q(x_t\mid x_{t-1}, y_t)$ and $\nu_t^{_\nii}$ such that they are independent of $\param$. By such a choice, we note that all the random elements in Algorithm~\ref{alg:pf}, $\pss{x^{_\nii}_{0:T},a^{_\nii}_{1:T}}$, become independent of $\param$. If we then condition on a certain realization of $\pss{x^{_\nii}_{0:T},a^{_\nii}_{1:T}}$, the estimate $\likest$ becomes a deterministic function in $\param$, and any standard optimization routine can subsequently be applied to solve~\eqref{eq:lik} and find $\MLparam$.

The strength of the particle filter, however, lies in the sequential build-up of the samples on the high-dimensional space $\mathsf{X}^{{T+1}}$, where the resampling operation provides important `feedback' on which parts of the state space to be explored further. With an arbitrary choice of $\param$-independent resampling weights $\nu_t^{_\nii}$, this feature will be lost, and we may expect an extremely high variance in the obtained estimate. In fact, a particle filter with $\param$-independent resampling weights $\nu_t^{_\nii}$ van be understood as importance sampling on the space $\mathsf{X}^{{T+1}}$, and we can in general not expect such an approach to be successful.

In order to obtain a deterministic function in $\param$, but avoid the bad consequences of a $\param$-independent resampling, we propose to let the resampling weights $\nu_t^{_\nii}$ and proposal $q(x_t\mid x_{t-1}, y_t)$ depend on some current parameter estimate $\theta_{k-1}$, as, e.g.,
\begin{subequations}\label{eq:nprw}
	\begin{equation}
	q(x_t\mid x_{t-1}, y_t) = f_{\param_{k-1}}(x_t\mid y_t)
	\end{equation} and \begin{equation}
	\nu_t^\nii = g_{\param_{k-1}}(y_t\mid x_t^\nii),
	\end{equation}
\end{subequations}
i.e., the bootstrap choice for $\param_{k-1}$ (instead of $\param$). If then $\param$ is somewhat close to $\param_{k-1}$, we can expect the variance of the corresponding estimate of the likelihood function, which we denote by $\likestn{\param_{k-1}}$, not to be forbiddingly large.

\begin{figure}[h]
	\centering\footnotesize
	\includegraphics{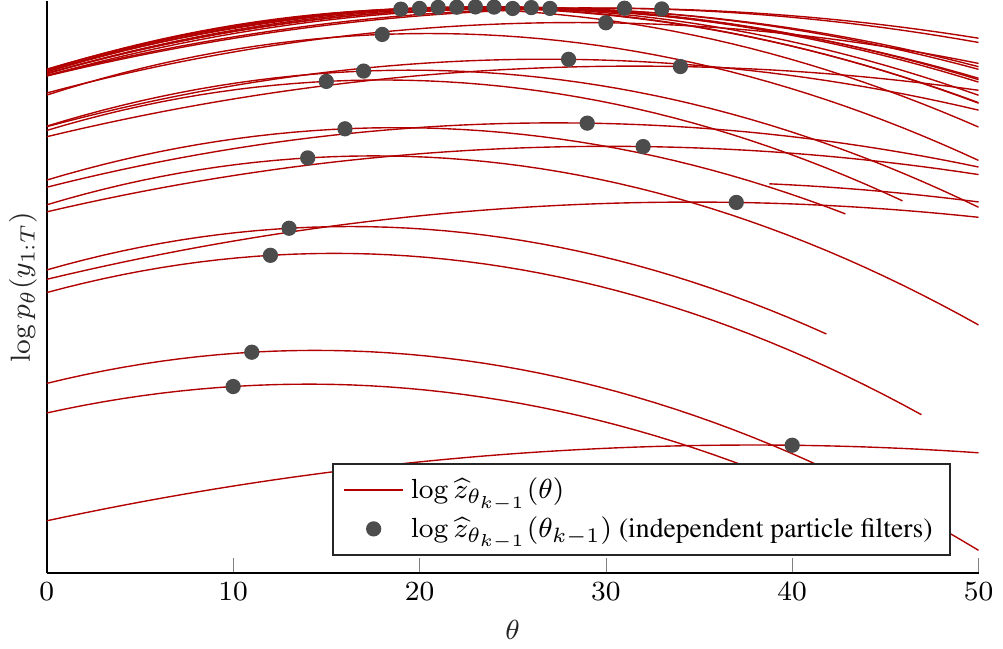}
	\caption{ The gray dots are log likelihood estimates obtained by running individual bootstrap particle filters with $N=100$ particles and parameter value $\theta_{k-1}$. Conditioned on the particle system underlying each particle filter (gray dots), the likelihood function is approximated also for other $\param$ values (red lines), which we can expect to be practically useful in the vicinity of $\param_{k-1}$. The idea of Algorithm~\ref{alg:idea} is as follows: Start with some $\theta_{k-1}$ and sample a corresponding gray dot (\texttt{particle\_filter}), and then apply a standard optimization scheme to find the maximum of the corresponding red line (\texttt{log\_likelihood}). We save the result as $\param_k$, and start over again with a new particle filter for~$\param_k$, etc.}
	\label{fig:idea}
\end{figure}

However, if the current $\param_{k-1}$ is far from $\MLparam$, we cannot expect $\likestn{\param_{k-1}}$ to be a particularly good estimator at the value $\MLparam$, and in particular, not expect the maximum of $\likestn{\param_{k-1}}$ to lie at $\MLparam$. For this reason, we have to iterate the parameter values over $k$ until we arrive in the vicinity of $\MLparam$. By inserting~\eqref{eq:nprw} into Algorithm~\ref{alg:pf}, combined with an outer optimization loop as discussed, and re-arranging, we arrive at Algorithm~\ref{alg:pfrr2}. For numerical reasons, we work with the logarithm of the likelihood function. The conceptual idea is illustrated in Figure~\ref{fig:idea}.

\begin{algorithm}[t]
	\SetKwFunction{prar}{particle\_filter}
	\SetKwFunction{likf}{log\_likelihood}
	Set $\param_{0}$\;
	\For{$k  = 1, \dots$}{
		Call $\pss{x^\nii_{0:T},a^\nii_{1:T}}\leftarrow$ \prar{$\param_{k-1}$}\;
		Solve \mbox{$\param_k \leftarrow \argmax_\param$ \likf{$\param,\param_{k-1},\pss{x^\nii_{0:T},a^\nii_{1:T}}$}}
		using an off-the-shelf optimization routine.\label{alg:pfrr2:max}\;
	}
	\BlankLine
	\setcounter{AlgoLine}{0}
	\SetKwProg{myprocp}{Function}{}{}
	\myprocp{\prar{$\param_{k-1}$}}{
		Draw ${x}_0^\nii \sim p({x}_0)$ and set $w_0^\nii = 1$.\;
		\For{$t = 1$ \KwTo $T$}{
			Draw ${a_t^\nii} \sim \mathcal{C}(\{w_{t-1}^\nij\}_{j=1}^N)$.\;
			Propagate ${x^{i}_{t}} \sim f_{\param_{k-1}}({x_{t}} \mid {x_{t-1}^{a_t^\nii}}, y_t)$.\;
			Set	$w_t^\nii \leftarrow g_{\theta_{k-1}}(y_t \mid x_t^\nii)$.\;
		}
		\KwRet $\pss{x^\nii_{0:T},a^\nii_{1:T}}$
	}
	\BlankLine
	\setcounter{AlgoLine}{0}
	\SetKwProg{myprocl}{Function}{}{}
	\myprocl{\likf{$\param,\param_{k-1},\pss{x^\nii_{0:T},a^\nii_{1:T}}$}}{
		\For{$t = 1$ \KwTo $T$}{
			Set $w_t^i \leftarrow \frac{w_{t-1}^{a_{t}^\nii}/\sum_j w_{t-1}^{a_{t}^\nij}}{(\star)}~\frac{f_{\param}(x_{t}|{x}_{t-1}^{a_{t}^\nii})}{f_{\param_{k-1}}(x_{t}|{x}_{t-1}^{a_{t}^\nii})} ~g_{\param}({y}_t|{x}_t^\nii)$.\;
			Set $z_t \leftarrow \frac{1}{N}\sum_{n=1}^{N}w_t^\nii$.\;
		}
		\KwRet $\log \likestn{\param_{k-1}} \leftarrow \sum_{t=1}^{T}\log z_t$.\;
	}
	\mycomment{$(\star) = g_{\param_{k-1}}(x_{t-1}^{a_{t}^\nii}\mid y_{t-1})/\sum_j g_{\param_{k-1}}(x_{t-1}^{a_{t}^\nij}\mid y_{t-1})$}
	\caption{Proposed method}
	\label{alg:pfrr2}
\end{algorithm}

\subsection{Solving the maximization problem}\label{sec:max}

On line~\ref{alg:pfrr2:max} in Algorithm~\ref{alg:pfrr2}, we have formulated a step where $\argmax_\param \likestn{\param_{k-1}}$ is to be solved. Importantly, this is now a completely deterministic problem and it can be handled by any standard numerical optimization tool. We will in the experiments demonstrate this by simply applying the general-purpose optimization tool \texttt{fminunc} in Matlab\footnote{Similar functions in other languages are \texttt{fminunc} (Octave), \texttt{scipy.optimize} (Python), \texttt{optim} (R) and \texttt{optimize} (Julia).}.




The particular structure of $\likestn{\param_{k-1}}$ (defined implicitly in the function \texttt{likelihood} in Algorithm~\ref{alg:pfrr2}) could, however, possibly be utilized by a more tailored optimization scheme. Its structure can be written as
\begin{align}\small
\likestn{\param_{k-1}} = \frac{1}{N}\prod_{t=1}^T\sum_{n=1}^{N}c_t^\nii~\omega_t^\nii(\param)~f_{\param}(x_{t}|{x}_{t-1}^{a_{t}^\nii})g_{\param}({y}_t|{x}_t^\nii),
\end{align}
where $c_t^\nii$ is a constant that is independent of $\param$, $\omega_t^{_\nii}(\param)$ depends on $\param$ but always fulfil $\sum_{n=1}^{N} \omega_t^{_\nii}(\param) = 1$, and $f_{\param}$ and $g_{\param}$ depends on the model. Whether this function exhibits any particular properties that could be exploited in optimization is subject to further research.


\section{Analysis}
We will now present a brief analysis of the proposed Algorithm~\ref{alg:pfrr2}. First, we conclude in Section~\ref{sec:a:conN} that the proposed scheme has desirable asymptotic properties as $N\to\infty$. Second, we make an attempt in Section~\ref{sec:a:conk} to understand the behavior also for finite $N$, and third in Section~\ref{sec:a:sgd} we discuss an alternative solution that would place the proposed method in the framework of stochastic gradient descent methods.

\subsection{Convergence as $N\to\infty$}\label{sec:a:conN}
Asymptotically as $N\to \infty$ in the particle filter, the proposed method becomes exact and converges (in principle) in one iteration. This can be realized as follows:
The log-likelihood is estimated by
\begin{align}
\log \lik \approx \sum_{t=1}^T \log\left( \sum_{i=1}^{N} w_t^\nii p(y_t\mid x_t^\nii)\right).\label{eq:MCapprox}
\end{align}
Assuming that the proposal distribution used to generate the particles $\{x_t^\nii\}_{t=1}^T$ is everywhere non-zero, this log-likelihood approximation is consistent under weak conditions and converges point-wise in $\theta$ as $N\rightarrow\infty$ \citep{DJ:2011,DelMoral:2004}. Thus, as long as the global solution to $\argmax_\param$ can be found, it is indeed the likelihood that has been maximized and $\MLparam$ found, which happens in a single iteration $k = 1$.

\subsection{Convergence as $k\to\infty$ and finite $N$}\label{sec:a:conk}

It is indeed reassuring that our proposed scheme is consistent as the number of particles $N\to\infty$, as discussed in the previous section. For practical purposes, however, the behavior for a finite $N$ (which always is the case in an implementation) is probably more interesting.

We start by noting that for $N<\infty$, it holds that $\Exp{\likestn{\param_{k-1}}} = \lik$ \citep{DelMoral:2004}. Note, however, that this does \emph{not} imply
\begin{align}
\Exp{\argmax_\param \likestn{\param_{k-1}}} = \argmax_\param p_\param(y),
\end{align}
so we do not have a theoretical justification to simply average the obtained sequence $\{\param_k\}$ to obtain $\MLparam$.

To obtain some guidance on how to extract a final estimate from the obtained sequence $\{\param_k\}$, we can make the simplifying assumption that the error $\log\likestn{\param_{k-1}}-\log \lik$, viewed as a stochastic process with index variable $\theta$, is stationary. In such a case, we can (under some technical assumptions) expect that $\MLparam$ is equal to the maximum mode of the distribution for $\param_k = \argmax_\param \log \likestn{\param_{k-1}}$. A proof sketch for this claim is found in Appendix~\ref{app:proof}. This suggests that we should look at the maximum mode in the histogram for $\{\param_k\}$ to find a good estimate of $\MLparam$ when we are using the method in practice (i.e., with a finite $N$). This will be illustrated in Example~2 and Figure~\ref{fig:ex2:hist}.

\subsection{Stability}
Stability, i.e., that the sequence $\param_0, \param_1, \dots$ does not diverge, is another important property. We have not experienced any such issues with the proposed scheme. The key to a stable algorithm is that the solution to the maximization problem on line~\ref{alg:pfrr2:max} in Algorithm~\ref{alg:pfrr2} is not too far from $\theta_{k-1}$. To motivate that this is likely to be the case, we note that while $\likestn{\theta}$ is unbiased for all $\param$, its variance $\sigma^2(\param)$ tends to increase as the distance between $\theta$ and $\theta_{k-1}$ increases. It is also known that $\log \likestn{\theta}$ (rather than $\likestn{\theta}$) has approximately a Gaussian distribution, which implies that $\log \likestn{\theta}$ has a bias in the order of $-\sigma^2(\param)$. This motivates that for $\param$ far from $\param_{k-1}$, the value of $\log \likestn{\theta}$ is likely to be small, and hence cause the algorithm not to deviate to much from the current iterate.

\subsection{Stochastic gradient descent}\label{sec:a:sgd}

An alternative approach would be not to solve the $\argmax_\param$ problem, but only use $\likestn{\param_{k-1}}$ to estimate a gradient around $\param_{k-1}$ and take an (inevitable stochastic) gradient step. Indeed, this has already been proposed by \citet{DT:2003}. Designing step lengths based on stochastic approximation ideas \citep{RM:1951} yields the well-studied stochastic gradient descent method. Our practical experience, however, is that (stochastic) gradient steps have inferior performance compared to the proposed $\argmax_\param$ scheme for our problem, including slower convergence and severe stability issues.

\section{Numerical experiments}
We will in this section apply our proposed method to two simulated examples, in order to illustrate and evaluate it. First a common example form the literature will be considered, and comparisons to alternative methods made. Thereafter a more difficult example will be studied. The code for the examples is available via the first author's homepage.

\subsection{Example 1}\label{ex:1}
\begin{figure}[t]
	\centering
	\begin{subfigure}{.49\linewidth}
		\includegraphics{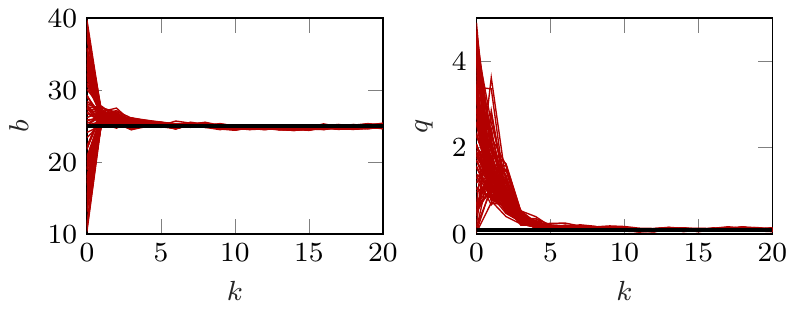}
		\caption{The proposed method.}
		\label{fig:ex1}
	\end{subfigure}
	\begin{subfigure}{.49\linewidth}
		\centering\footnotesize
		\includegraphics{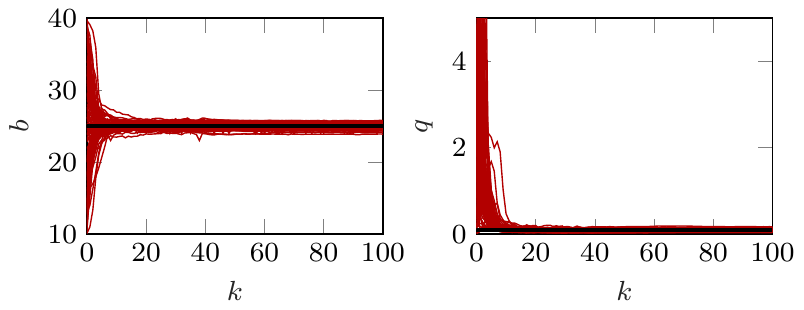}
		\vspace*{-1em}
		\caption{PSAEM \citep{Lindsten:2013}.}
		\label{fig:ex1:apsaem}
	\end{subfigure}	

~

~

	\begin{subfigure}{.49\linewidth}
		\centering\footnotesize
		\includegraphics{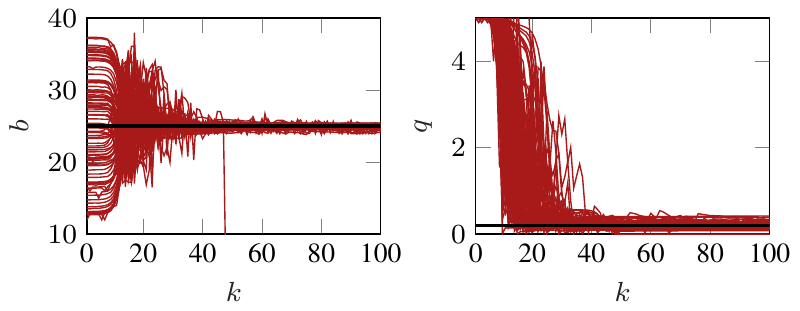}
		\vspace*{-1em}
		\caption{GP-based optimization \citep{WS:2017}.}
		\label{fig:ex1:GP}
	\end{subfigure}	
	\caption{ Example 1. The results from each method is shown by red lines, and the true parameter values are shown in black. The practical convergence properties of the proposed method appears to be promising.}
\end{figure}

In this first example, we consider $T=100$ measurements generated by the model
\begin{subequations}
	\begin{align}
	x_{t+1} &= 0.5x_t + b\tfrac{x_t}{1+x_t^2} + 8\cos(1.2t) + qw_t,\\
	y_t &= 0.05x_t^2+e_t,
	\end{align}
\end{subequations}
where $w_t\sim\mathcal{N}(0,1)$, $e_t\sim\mathcal{N}(0,1)$, and $\param=\{b,q\}$. The true values of $\param$ are $\{25,\sqrt{0.1}\}$, and this example (with $q = \sqrt{0.1}$ and $\theta = b$) was also used to generate Figure~\ref{fig:idea}. The proposed Algorithm~\ref{alg:pfrr2} is implemented with $N=100$, and employing the generic optimization routine \texttt{fminunc} in Matlab to solve the optimization problem on line~\ref{alg:pfrr2:max} in Algorithm~\ref{alg:pfrr2}. The initial~$\param_0$ is chosen randomly on the intervals $[10,40]$ and $(0,4]$, respectively, and the entire example is repeated 100 times. Each example took approximately 6 seconds on a standard laptop, and the results are found in Figure~\ref{fig:ex1}. We compare with two alternative methods for maximum likelihood estimation in nonlinear state-space models: The results for particle stochastic approximation EM (PSAEM, \citealt{Lindsten:2013}) applied to the very same problem are reported in Figure~\ref{fig:ex1:apsaem}. The results for the same problem with a stochastic optimization approach using the particle filter to estimate the likelihood and the Gaussian process (GP) to model the likelihood surface and its derivatives \citep{WS:2017,MH:2015} are found in Figure~\ref{fig:ex1:GP}. With the number of iterations chosen as in the figures, the computational load are of the same order of magnitude for all three methods.

From this, we conclude that our proposed method tend to converge faster than the alternatives (counting the number of iterations needed), but that each iteration is computationally more involved. The computational load of our algorithm is partly due to the use of a generic optimization routine (\texttt{fminunc} in Matlab), which makes no use of the particular structure of the objective function (\texttt{likelihood} in Algorithm~\ref{alg:pfrr2}), as discussed in Section~\ref{sec:max}.

\subsection{Example 2}
\begin{figure}[t]
	\centering
	\begin{subfigure}{.6\linewidth}
		\centering\footnotesize
		\includegraphics{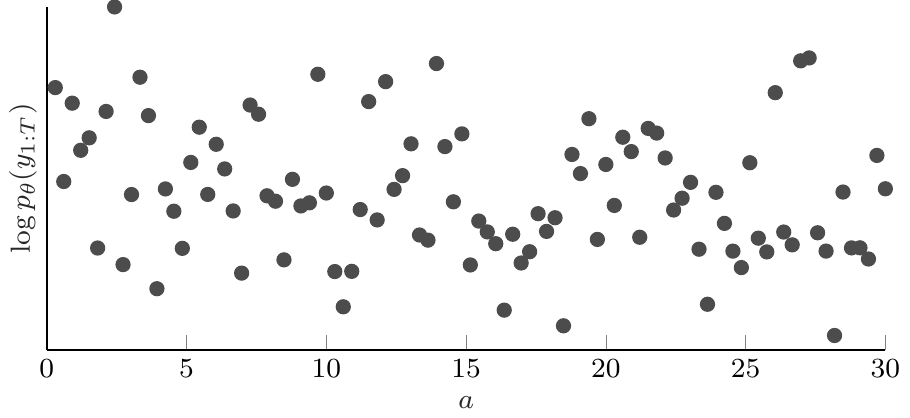}
		\caption{Log-likelihood estimates (vertical axis) for the model~\eqref{eq:ex:2:f} in Example~2 for different $a$ (horizontal axis, true $a=0.5$) and $b = -2$, obtained with independent particle filters with $N = 100$. As can be seen, the variance in $\likest$ is rather high in this example, which is to be compared with the gray dots in Figure~\ref{fig:idea}, the corresponding plot for Example~1. We thus expect maximum likelihood estimation to be significantly more challenging in this example.}
		\label{fig:ex2:loglik}
	\end{subfigure}
	
	~
	
	~
	
	\begin{subfigure}{.49\linewidth}
		\centering\footnotesize
		\includegraphics{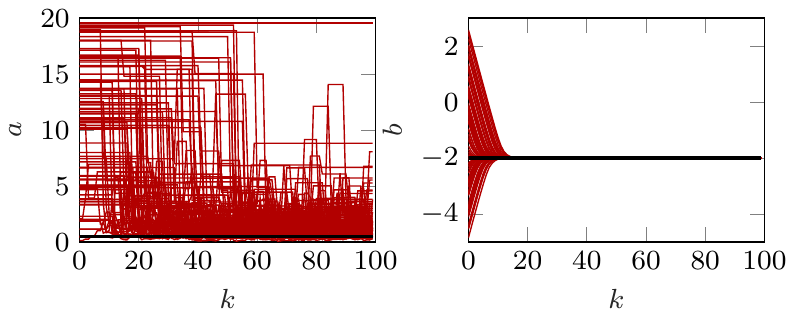}
		\caption{ The results for our proposed method on Example~2. Our proposed method manages, despite the poor quality likelihood estimates (Figure~\ref{fig:ex2:loglik}), to eventually converge towards sensible parameter values for a wide range of initializations. These traces are shown as histograms in the figure below.}
		\label{fig:ex2}
	\end{subfigure}\hfill
	\begin{subfigure}{.49\linewidth}
		\centering\footnotesize
		\includegraphics{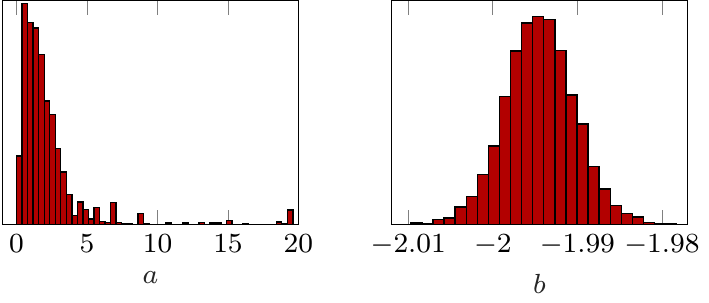}
		\caption{ The traces from Figure~\ref{fig:ex2} above as a histograms, after discarding the transient phase up to $k=50$. Using the principle suggested in Section~\ref{sec:a:conk}, the final estimate~$\MLparam$ should be taken as the maximum of the histograms, i.e., ${0.59, -1.995}$ , which corresponds well to the true parameter values $\{0.5,-2\}$.}
		\label{fig:ex2:hist}
	\end{subfigure}
	\caption{ Results for Example 2.}\label{fig:ex2:g}
\end{figure}

Now, consider the following state-space model
\begin{subequations}\label{eq:ex:2}
	\begin{alignat}{3}
	x_{t+1} &= \frac{x}{a + x^2} + b u + w_t,& ~~ w_t&\sim \N(0,1),\label{eq:ex:2:f}\\
	y_t &= x + e_t,& ~~ e_t&\sim \N(0,1).
	\end{alignat}
\end{subequations}
with $T = 1\,000$ and $\param = \{a, b\}$. One data set is generated with $\param = \{0.5, -2\}$, and our method is applied, with different initializations, 100 times to find $\MLparam$. This problem is significantly harder than Example~1 due to the location of $a$ in the denominator (and not the numerator) in~\eqref{eq:ex:2:f}. As an illustration, independent  particles filter were run to estimate the log-likelihood for different values of $a$ in Figure~\ref{fig:ex2:loglik}, from which we conclude that the likelihood estimate is rather noisy. This can be compared to Example~1 and the gray dots in Figure~\ref{fig:idea}, where the likelihood estimation is clearly less challenging. Again, we use the proposed Algorithm~\ref{alg:pfrr2} with \texttt{fminunc} in Matlab to solve the optimization problem on line~\ref{alg:pfrr2:max}. The results are shown in Figure~\ref{fig:ex2} and \ref{fig:ex2:hist}. Despite the challenging likelihood estimation, our proposed method manages to eventually converge towards meaningful values, and following the guidance discussed in Section~\ref{sec:a:conk}, we take the final estimates as the maximum of the histograms in Figure~\ref{fig:ex2:hist}, ${0.59, -1.995}$, which corresponds well to the true parameters.

\section{Conclusions}

We have proposed a novel method, Algorithm~\ref{alg:pfrr2}, to find maximum likelihood estimates of unknown parameters in nonlinear state-space models. The method builds on the particle filter, and allows for parameter inference in any model where also a particle filter can be applied. In fact, the method can be understood as a particular choice of $\param$-independent proposal and resampling weights in the auxiliary particle filter.

One key reason for the promising results is probably that we propose to solve an optimization problem at each iteration, instead of only taking a gradient step or similar: heuristically this seems to lead to fast convergence, and has not caused any problems with instability. The theoretical analysis, however, becomes more involved. We have presented an attempt to such an analysis in Section~\ref{sec:a:conk}, but all questions have not been answered. As mentioned, the work by \citet{LeGland:2007} could potentially be useful in a more thorough analysis of the proposed method.

A tailored choice of optimization routine would be interesting for further work. Furthermore, the applicability of the proposed method for the particle marginal Metropolis-Hastings sampler \citep{ADH:2010} would be another interesting question.

\bibliography{../../references}

\begin{thebibliography}{21}
\providecommand{\natexlab}[1]{#1}
\providecommand{\url}[1]{\texttt{#1}}
\expandafter\ifx\csname urlstyle\endcsname\relax
  \providecommand{\doi}[1]{doi: #1}\else
  \providecommand{\doi}{doi: \begingroup \urlstyle{rm}\Url}\fi

\bibitem[Andrieu et~al.(2010)Andrieu, Doucet, and Holenstein]{ADH:2010}
Christophe Andrieu, Arnaud Doucet, and Roman Holenstein.
\newblock Particle {Markov} chain {Monte} {Carlo} methods.
\newblock \emph{Journal of the Royal Statistical Society: Series B (Statistical
  Methodology)}, 72\penalty0 (3):\penalty0 269--342, 2010.

\bibitem[Dahlin and Lindsten(2014)]{DL:2014}
Johan Dahlin and Fredrik Lindsten.
\newblock Particle filter-based {Gaussian} process optimisation for parameter
  inference.
\newblock In \emph{Proceedings of the 19\textsuperscript{th} IFAC World
  Congress}, pages 8675--8680, Cape Town, South Africa, August 2014.

\bibitem[Del~Moral(2004)]{DelMoral:2004}
Pierre Del~Moral.
\newblock \emph{{Feynman}-{Kac} formulae: genealogical and interacting particle
  systems with applications.}
\newblock Springer, New York, NY, US, 2004.

\bibitem[Doucet and Johansen(2011)]{DJ:2011}
Arnaud Doucet and Adam~M. Johansen.
\newblock A tutorial on particle filtering and smoothing: fifteen years later.
\newblock In D.~Crisan and B.~Rozovsky, editors, \emph{Nonlinear Filtering
  Handbook}, pages 656--704. Oxford University Press, Oxford, UK, 2011.

\bibitem[Doucet and Tadi\'{c}(2003)]{DT:2003}
Arnaud Doucet and Vladislav~B. Tadi\'{c}.
\newblock Parameter estimation in general state-space models using particle
  methods.
\newblock \emph{Annals of the Institute of Statistical Mathematics},
  55\penalty0 (2):\penalty0 409--422, 2003.

\bibitem[Doucet et~al.(2000)Doucet, Godsill, and Andrieu]{DGA:2000}
Arnaud Doucet, Simon~J. Godsill, and Christophe Andrieu.
\newblock On sequential {Monte} {Carlo} sampling methods for {Bayesian}
  filtering.
\newblock \emph{Statistics and Computing}, 10\penalty0 (3):\penalty0 197--208,
  2000.

\bibitem[Doucet et~al.(2001)Doucet, de~Freitas, and Gordon]{DFG:2001}
Arnaud Doucet, Nando de~Freitas, and Neil~J. Gordon.
\newblock An introduction to sequential {Monte} {Carlo} methods.
\newblock In \emph{Sequential {Monte} {Carlo} methods in practice}, pages
  3--14. Springer, 2001.

\bibitem[Gordon et~al.(1993)Gordon, Salmond, and Smith]{GSS:1993}
Neil~J. Gordon, David~J. Salmond, and Adrian~F.M. Smith.
\newblock Novel approach to nonlinear/non-{Gaussian} {Bayesian} state
  estimation.
\newblock In \emph{IEE Proceedings F - Radar and Signal Processing}, pages
  107--113, 1993.

\bibitem[Kitagawa(1996)]{Kitagawa:1996}
Genshiro Kitagawa.
\newblock Monte carlo filter and smoother for non-gaussian nonlinear state
  space models.
\newblock \emph{Journal of Computational and Graphical Statistics}, 5\penalty0
  (1):\penalty0 1--25, 1996.

\bibitem[Le~Gland(2007)]{LeGland:2007}
Francois Le~Gland.
\newblock Combined use of importance weights and resampling weights in
  sequential monte carlo methods.
\newblock \emph{ESAIM: Proc.}, 19:\penalty0 85--100, 2007.

\bibitem[Lindsten(2013)]{Lindsten:2013}
Fredrik Lindsten.
\newblock An efficient stochastic approximation {EM} algorithm using
  conditional particle filters.
\newblock In \emph{Proceedings of the 38\textsuperscript{th} International
  Conference on Acoustics, Speech, and Signal Processing (ICASSP)}, pages
  6274--6278, Vancouver, BC, Canada, May 2013.

\bibitem[Mahsereci and Hennig(2015)]{MH:2015}
Maren Mahsereci and Philipp Hennig.
\newblock Probabilistic line searches for stochastic optimization.
\newblock In \emph{Advances in Neural Information Processing Systems 28
  (NIPS)}, pages 181--189, Montr\'{e}al, QC, Canada, December 2015.

\bibitem[Malik and Pitt(2011)]{MP:2011}
Sheheryar Malik and Michael~K. Pitt.
\newblock Particle filters for continuous likelihood evaluation and
  maximisation.
\newblock \emph{Journal of Econometrics}, 165\penalty0 (2):\penalty0 190--209,
  2011.

\bibitem[Naesseth et~al.(2015)Naesseth, Lindsten, and Sch\"{o}n]{NLS:2015}
Christian~A. Naesseth, Fredrik Lindsten, and Thomas~B. Sch\"{o}n.
\newblock Nested sequential {Monte} {Carlo} methods.
\newblock In \emph{Proceedings of the 32\textsuperscript{nd} International
  Conference on Machine Learning (ICML)}, pages 1292--1301, Lille, France, July
  2015.

\bibitem[Olsson et~al.(2008)Olsson, Capp{\'{e}}, Douc, and Moulines]{ODC+:2008}
Jimmy Olsson, Olivier Capp{\'{e}}, Randal Douc, and \'{E}ric Moulines.
\newblock Sequential {Monte} {Carlo} smoothing with application to parameter
  estimation in nonlinear state-space models.
\newblock \emph{Bernoulli}, 14\penalty0 (1):\penalty0 155--179, 2008.

\bibitem[Pitt and Shephard(1999)]{PS:1999}
Michael~K. Pitt and Neil Shephard.
\newblock Filtering via simulation: auxiliary particle filters.
\newblock \emph{Journal of the American Statistical Association}, 94\penalty0
  (446):\penalty0 590--599, 1999.

\bibitem[Robbins and Monro(1951)]{RM:1951}
Herbert Robbins and Sutton Monro.
\newblock A stochastic approximation method.
\newblock \emph{The Annals of Mathematical Statistics}, 22\penalty0
  (3):\penalty0 400--407, 1951.

\bibitem[Sch\"{o}n and Lindsten(2017)]{SL:2016}
Thomas~B. Sch\"{o}n and Fredrik Lindsten.
\newblock Sequential {Monte} {Carlo} methods.
\newblock 2017.

\bibitem[Sch\"{o}n et~al.(2011)Sch\"{o}n, Wills, and Ninness]{SWN:2011}
Thomas~B. Sch\"{o}n, Adrian Wills, and Brett Ninness.
\newblock System identification of nonlinear state-space models.
\newblock \emph{Automatica}, 47\penalty0 (1):\penalty0 39--49, 2011.

\bibitem[Svensson et~al.(2017)Svensson, Sch\"{o}n, and Lindsten]{SSL:2017}
Andreas Svensson, Thomas~B. Sch\"{o}n, and Fredrik Lindsten.
\newblock Learning of state-space models with highly informative observations:
  a tempered sequential monte carlo solution.
\newblock \emph{Mechanical Systems and Signal Processing}, 2017.
\newblock Accepted for publication.

\bibitem[Wills and Sch\"{o}n(2017)]{WS:2017}
Adrian Wills and Thomas~B. Sch\"{o}n.
\newblock On the construction of probabilistic newton-type algorithms.
\newblock In \emph{Proceedings of the 56\textsuperscript{th} IEEE Conference on
  Decision and Control (CDC)}, Melbourne, Australia, December 2017.

\end{thebibliography}

\appendix
\section{Proof sketch}\label{app:proof}

\small
This is a sketch for a proof of the claim given in Section~\ref{sec:a:conk}.
Let $\param_{k-1}$ be fixed and assume that $\varepsilon(\theta) = \log\likestn{\param_{k-1}} - \log\lik$ is a stationary stochastic process with index set $\theta \in \Theta$. For any $\theta'\in \Theta$ and $\delta > 0$, let $B_\delta(\theta')$ be a ball of radius $\delta$ centered at $\theta'$. For notational simplicity, let $h(\theta) = \log\lik$ and note that this is a deterministic function of $\theta$ which is assumed to be Lipschitz continuous and attain its maximum for $\theta=\MLparam$. Now, take $\delta$ sufficiently small so that $\min_{\theta\in B_\delta(\MLparam)} h(\theta) \geq \max_{\theta\notin B_\delta(\MLparam)} h(\theta)$. For any $\theta'$ with $\|\theta' - \MLparam\| \geq \delta$ we then have
\begin{multline*}
\Prb{\argmax_\theta \{\varepsilon(\theta) + h(\theta) \} \in B_\delta(\theta')} \\
= \Prb{\max_{\theta \in B_\delta(\theta')} \{\varepsilon(\theta) + h(\theta) \} \geq \max_{\theta\in\Theta} \{\varepsilon(\theta) + h(\theta) \} } \\
\leq \Prb{\max_{\theta \in B_\delta(\theta')} \varepsilon(\theta) + \min_{\rule{0pt}{2ex}\theta\in B_\delta(\MLparam)}h(\theta) \geq \max_{\theta\in\Theta} \{\varepsilon(\theta) + h(\theta) \} } \\
= \Prb{\max_{\rule{0pt}{2ex}\theta \in B_\delta(\MLparam)} \varepsilon(\theta) +  \min_{\rule{0pt}{2ex}\theta\in B_\delta(\MLparam)}h(\theta) \geq \max_{\theta\in\Theta} \{\varepsilon(\theta) + h(\theta) \} } \\
\leq \Prb{\max_{\rule{0pt}{2ex}\theta \in B_\delta(\MLparam)} \{\varepsilon(\theta) + h(\theta) \} \geq \max_{\theta\in\Theta} \{\varepsilon(\theta) + h(\theta) \} } \\
=\Prb{\argmax_\theta \{\varepsilon(\theta) + h(\theta) \} \in B_\delta(\MLparam)} \\
\end{multline*}
where the fourth line follows from the assumed stationarity of $\varepsilon(\theta)$. 
Now, since $\delta$ is arbitrary, it follows that if $X=\argmax_\theta \{\varepsilon(\theta) + h(\theta) \}$
admits a density w.r.t.\ Lebesgue measure, its density function $p_X(x)$ is maximized at $x=\MLparam$.

\end{document}